\PassOptionsToPackage{dvipsnames,svgnames,x11names}{xcolor}
\documentclass[11pt]{article}
\usepackage[a4paper, total={6.3in, 9.7in}]{geometry}
\usepackage{setspace} 
\usepackage{authblk}
\usepackage{graphicx,xcolor}
\usepackage{doi}
\usepackage{cite}
\usepackage{booktabs}
\usepackage{amsmath,mathtools,amssymb}
\usepackage{hyperref}

\usepackage{etoolbox}

\makeatletter
\patchcmd{\env@cases}{1.2}{0.72}{}{}
\makeatother

\hypersetup{
    colorlinks,
    linkcolor={red!50!black},
    citecolor={blue!50!black},
    urlcolor={blue!80!black}
}

\definecolor{LightGray}{gray}{0.9}

\DeclareMathOperator{\Tr}{Tr}

\title{\Large\bf 
Common origin of dark matter, baryon asymmetry and neutrino masses in the standard model with extended scalars
}
\author[1]{Sin Kyu Kang\footnote{\href{mailto:skkang@seoultech.ac.kr}{skkang@seoultech.ac.kr}}}
\author[2]{Raymundo Ramos\footnote{\href{mailto:rayramosang@gmail.com}{rayramosang@gmail.com}}}
\affil[1]{\textit{School of Natural Sciences, Seoultech,  Seoul 01811, Korea}}
\affil[2]{\textit{Institute of Convergence Fundamental Studies, Seoultech, Seoul 01811, Korea}}

\date{}

\begin{document}

\maketitle

\begin{abstract}
\noindent 
We propose a model that simultaneously addresses the existence of a dark matter candidate, baryon asymmetry and tiny neutrino masses and mixing by 
introducing two $SU(2)$ triplet scalars and an inert $SU(2)$ doublet scalar on top of the standard model.
The two triplet scalars serve as mediators in generation of lepton asymmetry and determination of relic density of dark matter. They also play an essential role in generation of tiny neutrino masses and inducing $CP$ violation.
The inert scalar is regarded as a dark matter candidate.
The interference due to complex Breit-Wigner propagators for the triplets
will result in $CP$-asymmetry that depends on the difference between their masses
and a relative complex phase between their couplings to standard model leptons.
Moreover, the production of lepton asymmetry
will be closely tied to the evolution of dark matter,
limiting the parameter space where the correct relic abundance and matter-antimatter asymmetry
can be simultaneously accomplished.
\end{abstract}

\newpage

% =======================

\section{Introduction}
\label{sec:intro}
Significant progress has been made in recent decades in accumulating evidence suggesting the presence of a mysterious, non-luminous form of matter known as dark matter (DM) in the current universe. Its quantity is estimated to be approximately five times greater than that of ordinary luminous or baryonic matter, denoted as $\Omega_B\simeq 5\%$~\cite{Planck:2018vyg}.
Among various new physics proposals for DM, the paradigm of weakly interacting massive particles (WIMPs) remains the most extensively studied scenario. In this scenario, a DM candidate, typically with a mass on the electroweak (EW) scale and interaction rates similar to EW interactions, can account for the correct DM relic abundance. 

On the other hand, the generation of baryon asymmetry~\cite{Sakharov:1967dj} through the out-of-equilibrium decay of a heavy particle has long been a well-established mechanism for baryogenesis~\cite{Weinberg:1979bt,Kolb:1979qa}. An intriguing approach to implement such a mechanism is through leptogenesis~\cite{Fukugita:1986hr}, where an initial net leptonic asymmetry is generated and subsequently converted into baryon asymmetry through $B + L$-violating EW sphaleron transitions. Remarkably, this scenario has the advantage of producing the necessary lepton asymmetry within the framework of the seesaw mechanism, which also provides an explanation for the origin of the minuscule neutrino masses~\cite{ParticleDataGroup:2022pth}. This represents yet another observed phenomenon that remains unaddressed by the Standard Model~(SM).

Recently, it was proposed that non-zero asymmetry
could also be generated by tree-level 2-to-2 scatterings mediated by unstable particles
with interfering diagrams~\cite{Dasgupta:2019lha}.
The main idea is to have $CP$-asymmetry generated by complex couplings
and decay widths in the propagators.
By taking two initial states to be DM and allowing the scattering process to violate lepton number, we can make the generation of baryon asymmetry via leptogenesis  intimately correlated with the DM relic density.

In this work, we take this idea to bring together a DM candidate and baryon asymmetry
by introducing an inert $SU(2)$ doublet (like in the inert Higgs doublet model(IDM)~\cite{Deshpande:1977rw,LopezHonorez:2006gr,Belyaev:2016lok,Gustafsson:2010zz}) and two $SU(2)$ triplet scalars that will have Yukawa couplings with a relative non-zero complex phase. We show that the tiny neutrino masses can be generated via
inverse type-II seesaw mechanism~\cite{Li:1985hy, Lusignoli:1990yk,deSPires:2005yok,Freitas:2014fda,deSousaPires:2018fnl}.
The inert doublet with $Z_2$ symmetry serves as a DM candidate and two triplets play
essential roles in generation of neutrino masses and lepton asymmetry that is correlated
with the DM relic density.
The numerical study conducted in this work illustrates that the model we propose can explain a DM candidate, baryon asymmetry and neutrino masses and mixing including $CP$ violation, simultaneously.

The rest of this paper is organized as follows:
in Sec.~\ref{sec:tripidm} we describe the main details of extending the SM with an inert doublet and two triplets,
in Sec.~\ref{sec:tripneu} we present the contributions from the Yukawa couplings between triplets and leptonic doublets,
in Sec.~\ref{sec:asymmetry} the origin of asymmetry in this model and its relation to DM is demonstrated,
while in Sec.~\ref{sec:numbers} we present the corresponding numerical results for successful benchmark points,
finally, in Sec.~\ref{sec:conclusion} we summarize and conclude.
Other relevant details about the scalar potential of the model are given in Appendix~\ref{app:scalars}.

\section{Two scalar triplets and an inert doublet in the standard model}
\label{sec:tripidm}

We propose new physics beyond the SM by introducing two $SU(2)$ scalar triplets
and an inert $SU(2)$ doublet,
together with an additional $Z_2$ symmetry.
Due to the odd charge of inert doublet under $Z_2$,
the dark sector is kept separated from the SM sector.
The additional scalar triplets will have even charges under the new $Z_2$ symmetry.
The triplets will also contribute to generate neutrino masses
via a non-zero vacuum expectation value (VEV).
The decomposition of the scalars of the model is as follows:
\begin{equation}
    \label{eq:scalars}
	\Phi_1 = \begin{pmatrix}
		0 \\
		\frac{1}{\sqrt{2}}(h_1 + v)
	\end{pmatrix},\quad
	\Phi_2 = \begin{pmatrix}
		H^+ \\
		\frac{1}{\sqrt{2}}(H^0 + i A^0)
	\end{pmatrix},\quad
	\Delta_{n} = \begin{pmatrix}
		\delta^+_{n}/\sqrt{2} & \delta^{++}_{n} \\
        \delta^0_{n} + u_n/\sqrt{2} & -\delta^+_{n}/\sqrt{2}
	\end{pmatrix},
\end{equation}
with $n\in \{1,2\}$.
More details about the scalar potential,
including minimization,
are given in Appendix~\ref{app:scalars}.
The VEVs of $\Phi_1$ and $\Delta_n$ are represented
by $v$ and $u_n$, respectively.
Following the notation of Ref.~\cite{Ferreira:2021bdj},
we will express the VEVs of $\Delta_1$ and $\Delta_2$
as $u_1 = u \cos\beta$ and $u_2 = u \sin\beta$, with $u = \sqrt{u_1^2 + u_2^2}$.
One part of the scalar potential,
of critical importance to the masses of neutrinos
and leptogenesis,
is the presence of trilinear terms between one doublet
and two triplets
\begin{equation}
    \label{eq:tripletpot}
    V \supset \sum_{n=1}^2 \left[ M_n^2 \Tr\left(\Delta_n^\dagger \Delta_n\right) +
    \left(\sum_{m=1}^2 \mu_{nm} \Phi_m^T i\sigma^2\Delta_n^\dagger \Phi_m + \text{H.c.}\right)\right],
\end{equation}
where the part between parenthesis breaks a global $U(1)$ symmetry in the potential.
Minimization of the potential relates $M_n^2$ with $\mu_{n1}$ as
\begin{eqnarray}
    M_1^2 \approx \frac{\mu_{11} v^2}{\sqrt{2} u \cos\beta},\quad
    M_2^2 \approx \frac{\mu_{21} v^2}{\sqrt{2} u \sin\beta}.
\label{Masses}
\end{eqnarray}
It is obvious from Eq.~(\ref{Masses}) that $u$ can be small when $M_n^2$ is large and/or $\mu_{n1}$ is small.
Taking  $\mu_{n1}$ to be small is technically natural in the sense that the global
$U(1)$ symmetry is recovered in the limit of
vanishing $\mu_{n1}$. 
This is an {\it inverse type-II seesaw} generating tiny neutrino masses.
The smallness of $u$ is responsible for tiny neutrino masses.
On the side of the couplings to $\Phi_2$, given by the terms
with $\mu_{n2}$,
we find that these terms correspond to trilinear vertices
that will appear in $s$-channel scatterings
communicating the dark sector with the leptonic sector of the SM,
with $\Delta_n$ as mediators.

\section{Leptonic couplings to the scalar triplets}
\label{sec:tripneu}

The scalar triplets can couple to left-handed lepton $SU(2)$ doublets, $L_j$.
We can write the following Yukawa terms
\begin{equation}
\label{eq:LagYuk}
	- \mathcal{L}_\mathrm{Yuk} =
		\sum_{n=1}^2 \sum_{j,k} Y^{\Delta_n}_{jk} L_j^T \mathcal{C}^\dagger i \tau_2 \Delta_n L_k + \mathrm{H.c.},
\end{equation}
where the indices $j$ and $k$ run over flavor indices $\{e,\mu,\tau\}$,
and $Y^{\Delta_n}$ can be understood as $3\times 3$ complex matrices of Yukawa couplings.
The charge conjugation matrix is represented by $\mathcal{C}$,
and $\tau_2$ is the second Pauli matrix.
Alas, this type of configuration
is known to result in flavor changing neutral currents~\cite{Pich:2009sp}.
To alleviate this effect, we can apply an alignment condition to the Yukawa couplings,
such that $Y^{\Delta_2} = \xi Y^{\Delta_1}$, with $\xi$ a complex coefficient.
For convenience, let us define $Y^\Delta \equiv Y^{\Delta_1} = Y^{\Delta_2} \xi^{-1}$.
Then, we can rewrite Eq.~\eqref{eq:LagYuk} as
\begin{equation}
\label{eq:LagYuk2}
  - \mathcal{L}_\mathrm{Yuk} =
    \sum_{j,k} Y^{\Delta}_{jk} L_j^T \mathcal{C}^\dagger i \tau_2 \left(
      \Delta_1 + \xi \Delta_2
    \right) L_k + \mathrm{H.c.}
\end{equation}

\subsection{Neutrino masses}
\label{sec:numass}

After the triplets $\Delta_n$ have acquired expectation values,
the mass terms can be read off from $\mathcal{L}_\mathrm{Yuk}$
and can be expressed as elements of a $3\times 3$ symmetric mass matrix
\begin{equation}
\label{eq:mnuyukawa}
  M^{\nu}_{jk} =
      \sqrt{2} Y^{\Delta}_{jk} u \cos\beta \left( 1 + \xi \tan\beta \right)\,.
\end{equation}
In the basis where charged lepton masses are diagonal,
this matrix can be diagonalized by the Pontecorvo-Maki-Nakagawa-Sakata (PMNS) matrix,
$U_\mathrm{PMNS}$
\begin{equation}
\label{eq:upmnsdiag}
	U_\mathrm{PMNS}^T M^\nu U_\mathrm{PMNS} = \mathrm{diag}(m^\nu_1, m^\nu_2, m^\nu_3) \equiv M^\nu_d.
\end{equation}
Using Eqs.~\eqref{eq:mnuyukawa} and~\eqref{eq:upmnsdiag},
we can use the measured oscillation parameters
to constrain $U_\mathrm{PMNS}$ and $M^\nu_d$
and relate them to the model parameters in $\mathcal{L}_\mathrm{Yuk}$
with
\begin{equation}
\label{eq:yukawa_upmns}
  \sqrt{2} Y^{\Delta}_{jk} u \cos\beta \left( 1 + \xi \tan\beta\right) =
    \left(U^*_\mathrm{PMNS} M^\nu_d U^\dagger_\mathrm{PMNS}\right)\,.
\end{equation}
The right hand side of the equation above
can be partially determined
from measured mixing angles,
$CP$-violating Dirac phase
and neutrino squared mass differences.
Additionally, two relative Majorana phases, $\phi_1$ and $\phi_2$,
need to be set to fully determine $U_\mathrm{PMNS}$.
To fully determine $M^\nu_d$, a value for the lightest neutrino mass
needs to be assumed.
Note that in Eq.~\eqref{eq:mnuyukawa},
assuming the factor $\sqrt{2} \cos\beta \left( 1 + \xi \tan\beta \right)$ is of $\mathcal{O}(1)$,
for the Yukawa couplings $Y^{\Delta}_{jk}$ to be also $\mathcal{O}(1)$
we would rely on $u$ to set the mass scale of the neutrinos.
With this argument we can expect $u$ to be below 1~eV.

\subsection{Couplings between leptons and scalars}
\label{sec:lepscouplings}

From Eq.~\eqref{eq:LagYuk}, we also obtain couplings
between components of the triplets and leptons.
Neutral scalar fields couple to pairs of neutrinos,
singly charged scalar fields couple to one charged lepton and one neutrino
and doubly charged fields couple to pairs of charged leptons.
Using the decomposition of the triplets shown in Eq.~\eqref{eq:scalars},
we can write such couplings from $\mathcal{L}_\mathrm{Yuk}$
\begin{equation}
    \mathcal{L}_\mathrm{Yuk} \supset
        Y^\Delta_{jk} \left[
            - \bar{\nu}^C_j \nu_k \left(\delta^0_1 + \xi \delta^0_2\right)
            + \sqrt{2} \bar{\nu}^C_j \ell_k \left(\delta^{+}_1 + \xi \delta^{+}_2\right)
            + \bar{\ell}^C_j \ell_k \left(\delta^{++}_1 + \xi \delta^{++}_2\right)
        \right]
        + \text{H.c.}
\end{equation}
These couplings, together with the trilinear couplings to $\Phi_2$,
contained in Eq.~\eqref{eq:tripletpot},
communicate the dark sector to the leptonic sector of the SM.
The fields in $\Delta_1$ and $\Delta_2$ work as mediators in these scatterings.
The way in which matter-antimatter asymmetry
can be achieved from unstable $\Delta_1$ and $\Delta_2$ with different masses,
and from the relative phase produced by $\xi$,
will be described in the next section.

\section{Origin of asymmetry}
\label{sec:asymmetry}

\begin{figure}[tb]
    \center
    \includegraphics[width=0.5\textwidth]{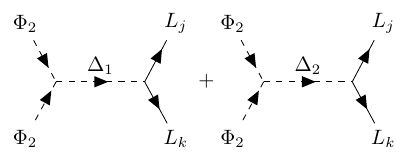}
    \caption{\label{fig:schan}%
        Feynman diagrams for processes mediated by $\Delta_n$
        with interference that contributes to matter asymmetry.
    }
\end{figure}

Following Ref.~\cite{Dasgupta:2019lha},
lepton asymmetry in this model is achieved
via interference effects between scatterings
mediated by unstable fields.
The case presented here corresponds
to interference between two $s$-channel processes,
labeled (i) in Sec.~II of said reference.
Consider the trilinear coupling of Eq.~\eqref{eq:tripletpot}
and the couplings between triplet and leptons in Eq.~\eqref{eq:LagYuk2}.
These couplings make possible the communication
between the dark sector and the SM leptonic sector,
via the Feynman diagrams displayed in Fig.~\ref{fig:schan}.
Two important elements to achieve asymmetry due to interference
are the complex Yukawa coupling,
most importantly through the relative phase contained in $\xi$,
and the presence of non-zero decay width in the Breit-Wigner propagators
of the triplets.
From Ref.~\cite{Dasgupta:2019lha} we have a $CP$-asymmetry factor
\begin{equation}
    \label{eq:cpasymfac}
    \delta \equiv |\mathcal{M}|^2 - |\bar{\mathcal{M}}|^2 =
        -4 \mathrm{Im}\left[\mathcal{C}_1\mathcal{C}_2^*\right]
        \mathrm{Im}\left[\mathcal{M}_1\mathcal{M}_2^*\right]|\mathcal{W}|^2\,,
\end{equation}
where we can identify
\begin{align}
    \label{eq:ImC1C2}
    \mathrm{Im}\left[\mathcal{C}_1\mathcal{C}_2^*\right] & =
        \left|Y^\Delta_{jk}\right|^2\mathrm{Im}\left[\mu_{12}\mu_{22}^*\xi^*\right]\,, \\
    \label{eq:ImM1M2}
    \mathrm{Im}\left[\mathcal{M}_1\mathcal{M}_2^*\right] & \propto
        \frac{
            \left(s - M_{\Delta_1}^2\right) M_{\Delta_2} \Gamma_{\Delta_2}
            - \left(s - M_{\Delta_2}^2\right) M_{\Delta_1} \Gamma_{\Delta_1}
        }{
            \left[\left(s - M_{\Delta_1}^2\right)^2 + M_{\Delta_1}^2\Gamma_{\Delta_1}^2\right]
            \left[\left(s - M_{\Delta_2}^2\right)^2 + M_{\Delta_2}^2\Gamma_{\Delta_2}^2\right]
        }\,,
\end{align}
and $\mathcal{W}$ contains wave functions for incoming and outgoing particles.
As mentioned before,
two critical characteristics
that can be readily identified from Eqs.~\eqref{eq:ImC1C2} and~\eqref{eq:ImM1M2},
are a non-zero $\mathrm{Im}\left[\mu_{12}\mu_{22}^*\xi^*\right]$%
---or at least for $\xi$ if we assumed the scalar potential parameters to be real---%
and non-zero decay widths for mediators.
Note also, that if the mediators shared the same masses and decay widths,
$\delta$ would vanish and, with it, asymmetry.
Moreover, this asymmetry is dependent on scatterings of the dark sector scalars
and, therefore, it will be affected by the evolution of DM.

\subsection{Lepton asymmetry and dark matter}
\label{sec:asymmdm}

As described in the last section,
the generation of asymmetry in the leptonic sector of the SM
requires 2-to-2 scatterings between dark sector particles and
SM leptons.
The contributing processes are
\begin{align}
\Phi_2^0+\Phi_2^0 \to \delta^0_n \to \nu_j + \nu_k\,, \\
H^+ + \Phi_2^0 \to \delta^+_n \to \ell_j + \nu_k\,, \\
H^+ + H^+ \to \delta^{++}_n \to \ell_j + \ell_k\,,
\end{align}
where $\Phi_2^0$ is used to represent neutral fields in $\Phi_2$.
In principle, these processes can contribute to the evolution of DM,
but we will find later that their contribution would be negligible.
The rest of the processes contributing to DM annihilation
are the usual annihilation to pairs of SM fermions
and to vector bosons $W^\pm$ and $Z$.
The Boltzmann equations that describe the evolution of DM
and leptonic asymmetry are
\begin{align}
        \label{eq:YPhi2eq}
    \frac{d Y_{\Phi_2}}{dx} = {} &
        \frac{-s}{H(x) x} \left(Y_{\Phi_2}^2 - Y_{\mathrm{eq},\Phi_2}^2\right)\langle \sigma v\rangle \left(\Phi_2\Phi_2 \to \mathrm{SM\,SM}\right)\,,\\
    \frac{d Y_{\Delta L}}{dx} = {} & \frac{s}{H(x) x} \bigl[
            \left(Y_{\Phi_2}^2 - Y_{\mathrm{eq},\Phi_2}^2\right)\langle \sigma v\rangle_\delta \left(\Phi_2\Phi_2 \to LL\right) \nonumber\\
        & - 2 Y_{\Delta L} Y_{\mathrm{eq},\Phi_2}^2 Y_{\mathrm{eq},\ell}^{-1}\langle \sigma v\rangle_\mathrm{tot} \left(\Phi_2\Phi_2 \to LL\right) \nonumber\\
        & - 2 Y_{\Delta L} Y_{\mathrm{eq},\Phi_2} \langle \sigma v\rangle_\mathrm{tot} \left(\Phi_2 \bar{L}\to\Phi_2^* L\right) \bigr]
    \label{eq:ydel}
\end{align}
where $x = m_\mathrm{L\Phi_2}/T$
with $m_\mathrm{L\Phi_2}$ standing for the lightest component of $\Phi_2$,
and $Y_{(eq,)k} = n_{(eq,)k}/s$ the (equilibrium) number densities normalized to entropy density, $s$,
for particle $k$.
The quantities $\langle \sigma v\rangle$ are thermally averaged cross sections times velocity.
We also used
\begin{equation}
    H(x) = \sqrt{\frac{8 \pi^3 g_*(T)}{90}} \frac{m_\mathrm{L\Phi_2}^2}{x^2 M_\mathrm{Pl}}
\end{equation}
where $g_*$ is the number of relativistic degrees of freedom at temperature $T$
and $M_\mathrm{Pl}$ is the Planck scale.
Necessarily, coannihilations have to be considered in all $\langle\sigma v\rangle$
and $\Phi_2$ stands for all the components taking part in the scatterings.
The averaged cross sections used in Eq.~\eqref{eq:ydel} are defined as
\begin{align}
    \label{eq:sigvdelta}
    \langle \sigma v\rangle_\delta \left(\Phi_2\Phi_2 \to LL\right) & \equiv
    \langle \sigma v\rangle \left(\Phi_2\Phi_2 \to LL\right)
    - \langle \sigma v\rangle \left(\Phi_2^*\Phi_2^* \to \bar{L}\bar{L}\right)\,, \\
    \label{eq:sigvtot}
    \langle \sigma v\rangle_\mathrm{tot} \left(\Phi_2\Phi_2 \to LL\right) & \equiv
    \langle \sigma v\rangle \left(\Phi_2\Phi_2 \to LL\right)
    + \langle \sigma v\rangle \left(\Phi_2^*\Phi_2^* \to \bar{L}\bar{L}\right)\,.
\end{align}
Solving Eqs.~\eqref{eq:YPhi2eq} and~\eqref{eq:ydel}
we obtain the evolution of the number density of DM
and the evolution of the leptonic asymmetry, $Y_{\Delta L}$.
Then we can convert the leptonic asymmetry to baryonic asymmetry, $Y_{\Delta B}$,
via the standard electroweak sphaleron process~\cite{Kuzmin:1985mm},
by considering the relationship $Y_{\Delta B} = -(28/51)Y_{\Delta L}$~\cite{Harvey:1990qw}
at the sphaleron temperature $T_\mathrm{sph} = 131.7 \pm 2.3$~GeV~\cite{DOnofrio:2014rug}.
Considering the requirement that the washout processes freeze-out
before DM do,
the most appropriate mass order for the lightest state of
$\Phi_2$ is above $\mathcal{O}(0.1)$~TeV,
where the main annihilation channel is to $W^\pm$ boson pair.
In this mass range, annihilation into fermions is subleading
and therefore expected to freeze-out before DM.
This puts DM mass in the region where
degeneracy is required to achieve the correct relic density.
Recent measurements put the relic density at $\Omega h^2 = 0.120\pm 0.001$,
and the observed baryon number asymmetry at $Y_{\Delta B} = (8.718 \pm 0.004)\times 10^{-11}$~\cite{Planck:2018vyg}.

\section{Numerical results}
\label{sec:numbers}

\begin{table}[tb]
    \center
    \setlength\tabcolsep{0.2cm}
    \begin{tabular}{lcccc}
        \toprule
        Parameter                  &   BP1    &   BP2  \\
        \cmidrule(r){1-1} \cmidrule(lr){2-3}
        $u$ [$10^{-11}$~GeV]       & 0.8598   & 1.129  \\
        $\tan\beta$                & 1.521    & 0.7533 \\
        $\mu_{11}$ [$10^{-9}$~GeV] & 1.317    & 4.391  \\
        $\mu_{21}$ [$10^{-9}$~GeV] & 1.421    & 2.231  \\
        $\mu_{12}$ [$10^{-1}$~GeV] & 1.534    & 2.650  \\
        $\mu_{22}$ [$10^{-1}$~GeV] & 1.868    & 3.255  \\
        $|\xi|$                    & 1.686    & 2.461  \\
        $\mathrm{ang}(\xi)$ [rad]  & 1.434    & 1.666  \\
        $\phi_{1}$ [rad]           & $-1.770$   & 0.8743 \\
        $\phi_{2}$ [rad]           & 0.5042   & 1.950  \\
        $m_{H^0}$ [TeV]             & 1.5      & 2.0    \\
        $m_{A^0}$ [TeV]             & 1.503    & 2.003  \\
        $M_{H^\pm}$ [TeV]          & 1.506    & 2.006  \\
        $\lambda_A$                & 0.21     & 0.3   \\
        $m_{\nu_1}$ [eV]           & \multicolumn{2}{c}{10$^{-3}$} \\
        $M_{12}^2$ [GeV$^2$]       &  \multicolumn{2}{c}{$(10^{-6})^2$} \\
        \bottomrule
    \end{tabular}
    \caption{\label{tab:benchmarkpts}%
        Numerical parameters for two benchmark points, labeled BP1 and BP2,
        used to solve Eqs.~\eqref{eq:YPhi2eq} and~\eqref{eq:ydel}.
        The corresponding solutions for $\Omega h^2$ and $Y_{\Delta B} = -(28/51) Y_{\Delta L}$
        are shown in Fig.~\ref{fig:BPevolution}.
        We have assumed normal hierarchy for the neutrino mass,
        i.e., $\nu_1$ is the lightest neutrino mass eigenstate.
        In our calculation we have used $\lambda_A = \lambda_{\Phi 12}  + \lambda'_{\Phi 12} - \lambda_5$.
    }
\end{table}

\begin{table}[tb]
    \center
    \setlength\tabcolsep{0.2cm}
    \begin{tabular}{lcccc}
        \toprule
                  & BP1 & BP2 \\
        \cmidrule(r){1-1} \cmidrule(lr){2-3}
        $Y^\Delta_{(1, 1)}$    & $0.1589 + i0.01129$  & $-0.0451 - i0.1547$ \\
        $Y^\Delta_{(1, 2)}$    & $-0.0145 + i0.3615$  & $0.0630 + i0.08962$ \\
        $Y^\Delta_{(1, 3)}$    & $-0.2073 + i0.1130$  & $ 0.1527 + i0.2377$ \\
        $Y^\Delta_{(2, 2)}$    & $0.358 - i1.348   $  & $-0.8976 - i0.8025$ \\
        $Y^\Delta_{(2, 3)}$    & $0.065 - i1.342   $  & $-0.6755 - i0.4886$ \\
        $Y^\Delta_{(3, 3)}$    & $0.2826 - i0.98834$  & $-0.6686 - i0.6317$ \\
        % \cmidrule(r){1-1} \cmidrule(lr){2-3}
        % $\Delta_n$ masses         & BP1 & BP2 \\
        % \cmidrule(r){1-1} \cmidrule(lr){2-3}
        $M_{\Delta_1}$ [GeV] & 3458 & 4569 \\
        $M_{\Delta_2}$ [GeV] & 2911 & 3753 \\
        % \cmidrule(r){1-1} \cmidrule(lr){2-3}
        % Direct detection         & BP1 & BP2 \\
        % \cmidrule(r){1-1} \cmidrule(lr){2-3}
        $\sigma_\mathrm{SI}$ [cm$^2$] & $6.984\times 10^{-46}$ & $8.02\times 10^{-46}$ \\
        \bottomrule
    \end{tabular}
    \caption{\label{tab:benchmarkptsres}%
        Relevant numerical results for the benchmark points of Table~\ref{tab:benchmarkpts}.
    }
\end{table}

\begin{figure}[tb]
    \center
    \includegraphics{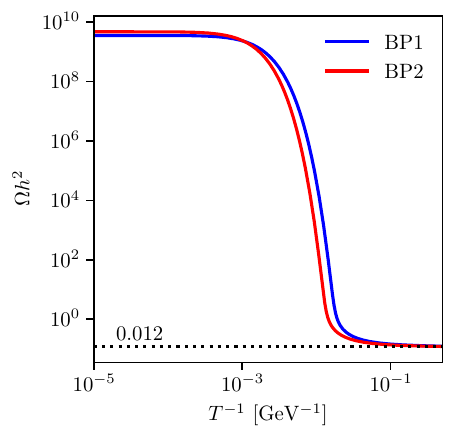}
    \includegraphics{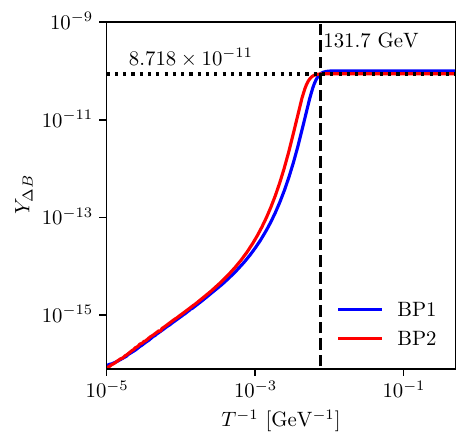}
    \caption{\label{fig:BPevolution}%
        Evolution of the relic density, $\Omega h^2$,
        and the asymmetry $Y_{\Delta B} = -(28/51) Y_{\Delta L}$,
        for the benchmark points given in Table~\ref{tab:benchmarkpts}.
        Horizontal dotted lines represent the central values for current measurements,
        $\Omega h^2 = 0.120\pm 0.001$ and $Y_{\Delta B} = (8.718 \pm 0.004)\times 10^{-11}$.
        The vertical dashed line is the central value for the sphaleron temperature,
        $T_\mathrm{sph} = 131.7 \pm 2.3$~GeV.
    }
\end{figure}

To test that it is possible to achieve correct relic density
and baryon asymmetry simultaneously
we implement the model presented in Secs.~\ref{sec:tripidm} and~\ref{sec:tripneu}
in \texttt{CalcHEP}~\cite{Belyaev:2012qa} to calculate squared amplitudes and decay widths.
These are used to calculate the averaged cross sections required in Eqs.~\eqref{eq:YPhi2eq}
and~\eqref{eq:ydel}.
Aformentioned equations are solved numerically
for two benchmark points
presented in Table~\ref{tab:benchmarkpts},
with the resulting evolution depicted in Fig.~\ref{fig:BPevolution}.

While the number of parameters present in the scalar potential of Appendix~\ref{app:scalars}
is enormous, in practice only a few parameters will actually have an important effect.
In the case of the parameters $\mu_{nm}$,
they are expected to have strong effects due to their relationship to the masses
of the components of $\Delta_1$ and $\Delta_2$ (for $\mu_{11}$ and $\mu_{21}$),
and the presence of $\mu_{12}$ and $\mu_{22}$ in Eq.~\eqref{eq:ImC1C2}.
From that same equation, it is obvious that a non-zero phase for $\xi$ will be necessary.
It is also expected that $u$ and $\beta$ play an important role
since they communicate the scalar sector and the neutrino masses.
In the case of the parameters of $U_\mathrm{PMNS}$,
we have taken the Majorana phases $\phi_1$ and $\phi_2$ as free parameters,
while $\sin^2\theta_{12,13,23}$ and the Dirac $CP$-violating phase have been fixed to their central
values given in the latest global fit by NuFIT~(5.2)~\cite{Gonzalez-Garcia:2021dve,nufitwebsite}.
To determine the values of $M^\nu_d$,
we fix the value of the lightest neutrino,
which we have chosen to be $\nu_1$ (normal hierarchy),
and obtain the other two using the central values of $\Delta m^2_{21}$
and $\Delta m^2_{31}$ from the same global fit.
One can consider the case of inverted hierarchy, but no significant change
in results is obtained.

One of the most notable features of the parameters in Table~\ref{tab:benchmarkpts},
is the size of $u$ ($10^{-11}$~GeV), the sizes of $\mu_{n1}$ ($10^{-9}$~GeV)
and the sizes of $\mu_{n2}$ ($10^{-1}$~GeV).
First of all, consider the processes depicted in Fig.~\ref{fig:schan},
with vertices proportional to the Yukawa couplings, $Y^{\Delta}_{jk}$,
and $\mu_{n2}$.
By inspection of Eq.~\eqref{eq:mnuyukawa},
achieving neutrino masses below $\mathcal{O}(1\ \text{eV})$
without strongly suppressing $Y^{\Delta}_{jk}$,
requires a small value for $u$ instead.
On the same footing, the values for $\mu_{n2}$ have to be
chosen in a mass scale where they allow for a sizable
contribution from the interference of the processes in Fig.~\ref{fig:schan}.
The resulting values for $Y^{\Delta}_{jk}$ are given in Table~\ref{tab:benchmarkptsres}.
Considering that $u$ has $\mathcal{O}(10^{-11}\ \text{GeV})$,
to avoid the suppression from massive $\Delta_n$
in the propagators of the same processes,
from Eqs.~\eqref{eq:mdelta1nomix} and~\eqref{eq:mdelta2nomix}
(or, equivalently, Eq.~\eqref{Masses})
we know that the two values of $\mu_{n1}$ have to be small as well.
In the case of the benchmark points of Table~\ref{tab:benchmarkptsres},
$\mu_{n1}$ of $\mathcal{O}(10^{-9}\ \text{GeV})$ results
in masses of $\mathcal{O}(1\ \text{TeV})$,
as shown in Table~\ref{tab:benchmarkptsres}.

\begin{figure}[tb]
    \center
    \includegraphics{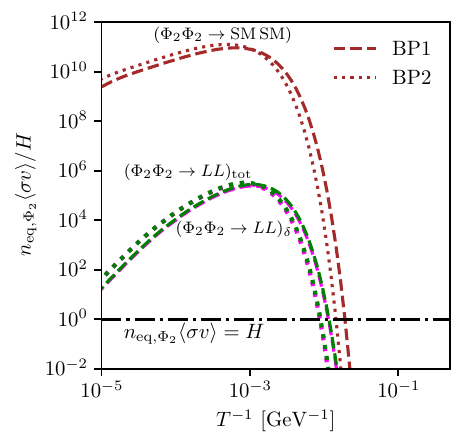}
    \caption{\label{fig:nsigmavh}%
        Interaction rates at equilibrium over Hubble parameter,
        $n_{\mathrm{eq},\Phi_2}\langle \sigma v \rangle/H$,
        for the two benchmark points (BP1: dashed, BP2: dotted).
        The brown lines correspond to the interaction $(\Phi_2 \Phi_2 \to \mathrm{SM\,SM})$,
        magenta lines are for the difference $(\Phi_2 \Phi_2 \to L L)_\delta$
        and green lines are for the total $(\Phi_2 \Phi_2 \to L L)_\mathrm{tot}$,
        with these last two interaction rates defined in Eqs.~\eqref{eq:sigvdelta} and~\eqref{eq:sigvtot}, respectively.
        Due to interference, the lines for $(\Phi_2 \Phi_2 \to L L)_\delta$ and $(\Phi_2 \Phi_2 \to L L)_\mathrm{tot}$
        are almost overlapping.
    }
\end{figure}

As can be seen in the left pane of Fig.~\ref{fig:BPevolution},
the evolution of DM relic density is quite standard.
This is expected from the form of Eq.~\eqref{eq:YPhi2eq},
which is standard by itself.
We can see that the onset of freeze-out of DM
is at a temperature around $T_f^{-1} \sim 10^{-2} \text{~GeV}^{-1}$,
which is consistent with $x_f \approx 27$ for the two benchmark points.
By looking at the brown taller lines in Fig.~\ref{fig:nsigmavh}
we see that this $x_f$ matches the expectation drawn from the size of $\langle \sigma v \rangle(\Phi_2\Phi_2 \to \text{SM\,SM})$.
On the right pane of Fig.~\ref{fig:BPevolution}
we show the evolution of $Y_{\Delta B} = -(28/51) Y_{\Delta L}$.
One notable feature of this figure
is the presence of an accelerated evolution
when approaching the sphaleron temperature, $T_\text{sph} = 131.7$~GeV.
This is due to the DM contribution in Eq.~\eqref{eq:ydel}
as it starts deviating from $Y_{\text{eq},\Phi_2}$
when approaching $x_f$.
For both benchmark points,
in Fig.~\ref{fig:nsigmavh} we can see
that the interaction rates to lepton pairs become smaller than $H$
before the cross section to pairs of all SM particles do.
This is convenient since we want the washout process
to fall behind the universe expansion before the freeze-out of DM.
% This is fine as long as the correct amount of $Y_{\Delta B}$ has been accumulated
% when $T = T_\text{sph}$.
Note that, in Fig.~\ref{fig:nsigmavh},
the $(\Phi_2\Phi_2 \to LL)$ and $(\Phi_2\Phi_2 \to \text{SM\,SM})$ interaction rates fall below $H$
for temperatures that are very close.
This ensures a larger contribution for $Y_{\Delta L}$
from DM falling out of chemical equilibrium.
Next, we comment on the interaction rates over Hubble parameter ($n_{eq,\Phi_2} \langle \sigma v \rangle/H$),
obtained with the benchmark points of Table~\ref{tab:benchmarkpts}
and displayed in Fig.~\ref{fig:nsigmavh}.
As pointed out in Sec.~\ref{sec:asymmdm},
the decay rate dominant for DM evolution, $\Phi_2 +\Phi_2 \to \text{SM} + \text{SM}$,
is several orders of magnitude larger than the washout process, $\Phi_2 +\Phi_2 \to L + L$.
Moreover, with the choice of $\mu_{nm}$ of Table~\ref{tab:benchmarkpts},
the $\langle \sigma v \rangle_\delta(\Phi_2\Phi_2 \to LL)$
has almost the same value as the sum of contributions $\langle \sigma v \rangle_\text{tot}(\Phi_2\Phi_2 \to L L)$.
This is due to the same interference effects that result in a non-zero $CP$-asymmetry in Eq.~\eqref{eq:cpasymfac},
and is what allows the large growth of $Y_{\Delta B}$, displayed in Fig.~\ref{fig:BPevolution}.

Lastly, the two main experimental probes on this model
would be comprised of search for the decays of the doubly charged
scalar and limits on direct detection of DM.
On the side of searches for the doubly charged scalar,
ATLAS and CMS report mass limits slightly below 1~Tev~\cite{CMS:2017pet,ATLAS:2017xqs}.
Having all the components of $\Delta_1$ and $\Delta_2$ with masses above TeV
puts them in a safe place for these searches.
In the case of direct detection of DM,
we use \texttt{micrOMEGAs}~\cite{Belanger:2020gnr} to calculate
spin independent direct detection cross sections, $\sigma_\text{SI}$,
and report the values in Table~\ref{tab:benchmarkptsres}.
By comparing against limits from the XENON1T experiment~\cite{XENON:2017vdw,XENON:2018voc},
we see that for DM masses of 1.5~TeV and 2~TeV,
the obtained values for $\sigma_\text{SI}$ are safely below exclusion limits.

\section{Conclusion}
\label{sec:conclusion}

Three of the most salient shortcomings of the SM are
matter asymmetry, the presence of massive neutrinos
and the existence of DM.
In this work we have demonstrated
that these three problems can be explained
with a scalar sector extended by two $SU(2)$ triplets and one inert doublet,
via the inverse type-II seesaw mechanism.
The two triplet scalars participate in producing light neutrino masses
via the type-II seesaw mechanism,
resulting in 2-to-2 scatterings that violate lepton number.
The use of an inert doublet includes an explanation for DM
while also aiding in the accumulation of lepton asymmetry that is
later converted to baryon asymmetry via the standard sphaleron process.
The accumulation of baryon asymmetry
is mostly controlled by the four possible trilinear couplings between two doublets
and one triplet.
On one hand, the couplings to to the SM-like doublet have to be quite small to avoid very massive
triplets.
On the other hand, the couplings to the inert doublet are required to be comparatively larger
to allow enough communication between the dark sector and the lepton number violating processes
that are responsible for matter asymmetry.
Similarly, the Yukawa couplings that also appear in said processes, if assumed to be $\mathcal{O}(1)$,
require a very small value for the VEVs of the triplets,
in order to achieve very light masses for the neutrinos.
For the benchmark points presented here,
direct detection cross section is well in reach of future experiments~\cite{LZ:2018qzl,PandaX:2018wtu,XENON:2020kmp,Billard:2021uyg},
while the masses of the triplets may still be above the reach of near-future colliders~\cite{deMelo:2019asm}.

\section*{Acknowledgments}

The work of R.R. was supported by the National Research Foundation of Korea under grant NRF-2021R1A2C4002551.

\appendix

\section{Extended scalar sector}
\label{app:scalars}

The scalar potential we consider
includes two $SU(2)_L$ triplets,
$\Delta_1$ and $\Delta_2$,
and one inert $SU(2)_L$ doublet, $\Phi_2$,
in addition to the Higgs doublet of the SM, $\Phi_1$.
The decomposition of the scalar fields is given explicitly in Eq.~(\ref{eq:scalars}).
%\begin{equation}
%	\Phi_1 = \begin{pmatrix}
%		0 \\
%		\frac{1}{\sqrt{2}}(v + h_1)
%	\end{pmatrix},\quad
%	%
%	\Phi_2 = \begin{pmatrix}
%		H^+ \\
%		\frac{1}{\sqrt{2}}(H^0 + i A^0)
%	\end{pmatrix},\quad
%	%
%	\Delta_{n} = \begin{pmatrix}
%		\delta^+_{n}/\sqrt{2} & \delta^{++}_{n} \\
%       \delta^0_{n} + u_n/\sqrt{2} & -\delta^+_{n}/\sqrt{2}
%	\end{pmatrix}
%\end{equation}
%with $n\in \{1,2\}$.
We also consider an additional
$Z_2$ symmetry
under which $\Phi_2$ is odd
while the rest of the fields are even.
The most general potential can be written as:
\begin{align}
	V = {}& V_\mathrm{IDM} + V_{\Delta}
		+ V_{H\Delta} + V_\mathrm{SB} \\
	V_\mathrm{IDM} = {}& - m_{\Phi 1}^2 \Phi_1^\dagger \Phi_1 + m_{\Phi 2}^2 \Phi_2^\dagger \Phi_2 + \lambda_{\Phi 1} (\Phi_1^\dagger \Phi_1)^2
		+ \lambda_{\Phi 2} (\Phi_2^\dagger \Phi_2)^2 \nonumber \\
		& + \lambda_{\Phi 12} \Phi_1^\dagger \Phi_1 \Phi_2^\dagger \Phi_2 + \lambda'_{\Phi 12} \Phi_1^\dagger \Phi_2 \Phi_2^\dagger \Phi_1
		  + \lambda_5 \mathrm{Re}\left[(\Phi_1^\dagger \Phi_2)^2\right] \\
	V_{\Delta} = {}& \sum_{n=1}^2 \left\{
            M_n^2 \Tr\left(\Delta_n^\dagger \Delta_n\right) + \lambda_{\Delta n} \Tr\left[\left(\Delta_n^\dagger \Delta_n\right)^2\right]
		 + \lambda'_{\Delta n} \left[\Tr\left(\Delta_n^\dagger \Delta_n\right)\right]^2\right\} \nonumber\\
		& + \lambda_{\Delta 12} \Tr(\Delta_1^\dagger \Delta_1 \Delta_2^\dagger \Delta_2)
		+ \lambda'_{\Delta 12} \Tr(\Delta_1^\dagger \Delta_1) \Tr(\Delta_2^\dagger \Delta_2) \nonumber\\
		& + \lambda_{\Delta 21} \Tr(\Delta_2^\dagger \Delta_1 \Delta_1^\dagger \Delta_2)
		+ \lambda'_{\Delta 21} \Tr(\Delta_2^\dagger \Delta_1) \Tr(\Delta_1^\dagger \Delta_2) \\
	V_{\Phi \Delta} = {}& \sum_{n=1}^{2}\sum_{k=1}^{2}\left[
            \lambda_{\Phi k\Delta n} \Phi_k^\dagger \Phi_k \Tr\left(\Delta_n^\dagger \Delta_n\right)
            + \lambda'_{\Phi k\Delta n} \Phi_k^\dagger \Delta_n  \Delta_n^\dagger \Phi_k
        \right] \\
	V_\text{SB} = {}&
        \sum_{m=1}^2 \left\{
            \sum_{n=1}^2 \mu_{nm} \Phi_m^T i\sigma^2\Delta_n^\dagger \Phi_m
            +  \lambda_{\Phi k\Delta 12} \Phi_m^\dagger \Phi_m \Tr\left(\Delta_1^\dagger \Delta_2 \right)
		+ \lambda'_{\Phi k\Delta 12} \Phi_m^\dagger \Delta_1 \Delta_2^\dagger \Phi_m\right\} \nonumber \\
        & M_{12}^2 \Tr\left(\Delta_1^\dagger \Delta_2\right)
        + \sum_{ijkl} \left[\lambda_{ijkl} \Tr\left(\Delta_i^\dagger \Delta_j \Delta_k^\dagger \Delta_l\right)
		+ \lambda'_{ijkl} \Tr\left(\Delta_i^\dagger \Delta_j\right)\left(\Delta_k^\dagger \Delta_l\right)\right] + \text{H.c.}
\end{align}
The terms in $V_{SB}$ break a global $U(1)$ symmetry.
The combination of indices $(i,j,k,l)$ can take the values
(2,1,1,1), (1,2,1,1), (1,2,2,2), (2,1,2,2) and (1,2,1,2);
other combinations belong to the H.c. part of the potential.
After electroweak symmetry breaking,
the scalar fields $\Phi_1$, $\Delta_1$ and $\Delta_2$
acquire VEVs of the form
\begin{equation}
	\langle \Phi_1\rangle = \frac{1}{\sqrt{2}}\begin{pmatrix}
		0 \\
		v
	\end{pmatrix},\quad
	\langle\Delta_{n}\rangle = \frac{1}{\sqrt{2}}\begin{pmatrix}
		0 & 0 \\
		u_n & 0
	\end{pmatrix}
\end{equation}
similarly to Ref.~\cite{Ferreira:2021bdj},
we will use $u_1^2 + u_2^2 = u^2$,
with $u_1 = u\cos\beta$, $u_2 = u\sin\beta$ and $\tan\beta = u_2/u_1$.
The VEVs must follow the condition $v^2 + 2u^2 \approx (246\ \text{GeV})^2$
which limits $u$ to be below 8~GeV due to constraints on the $\rho$ parameter.
From the first derivatives of the potential we can find the following conditions
\begin{align}
	\label{eq:cond1}
    0 = {}
		& \lambda_{\Phi 1} v^2 - m^2_{\Phi 1}
		+ u^2 \left(
			\lambda^q_{\Phi 1 \Delta 1} \cos^2\beta
			+ \lambda^q_{\Phi 1\Delta 2} \sin^2\beta
			+ 2 \lambda^q_{\Phi 1 \Delta 12} \cos\beta \sin\beta
		\right) \nonumber\\
		 & - \sqrt{2} u \left(\mu_{11} \cos\beta + \mu_{21} \sin\beta\right) \\
	\label{eq:cond2}
	0 = {}&
		u^3 \left[
			(\lambda^q_{12} + \lambda^q_{1212}) \cos\beta \sin^2\beta
			+ \lambda^q_1 \cos^3\beta
			+ 3 \left(\lambda^q_{2111} + \lambda^{q\prime}_{2111}\right) \cos^2\beta \sin\beta
			+ \left(\lambda^q_{1222} + \lambda^{q\prime}_{1222}\right) \sin^3\beta
			\right] \nonumber\\
				& + u \left[
					M^2_1 \cos\beta
					+ M^2_{12} \sin\beta
					+ \lambda^q_{\Phi 1\Delta 12} v^2 \sin\beta
					+ \lambda^q_{\Phi 1\Delta 1} v^2 \cos\beta
					\right]
					- \frac{\mu_{11}}{\sqrt{2}} v^2 \\
	\label{eq:cond3}
    0 = {}&
        u^3 \left[
            (\lambda^q_{12} + \lambda^q_{1212}) \cos^2\beta \sin\beta
            + \lambda^q_2 \sin^3\beta
            + 3 \left(\lambda^q_{1222} + \lambda^{q\prime}_{1222}\right) \cos\beta \sin^2\beta
            + \left(\lambda^q_{2111} + \lambda^{q\prime}_{2111}\right) \cos^3\beta
        \right] \nonumber\\
        & + u \left[
            M^2_2 \sin\beta
            + M^2_{12} \cos\beta
            + \lambda^q_{\Phi 1\Delta 12} v^2 \cos\beta
            + \lambda^q_{\Phi 1\Delta 2} v^2 \sin\beta
        \right]
        - \frac{\mu_{21}}{\sqrt{2}} v^2
\end{align}
where we used the following definitions
\begin{align}
	\lambda^q_n \equiv {}& \lambda_{\Delta n} + \lambda'_{\Delta n}, \\
	\lambda^q_{\Phi 1\Delta n} \equiv {}& (\lambda_{\Phi 1\Delta n} + \lambda'_{\Phi 1\Delta n})/2, \\
	\lambda^q_{12} \equiv {}& (\lambda_{\Delta 12} + \lambda_{\Delta 21} + \lambda'_{\Delta 12} + \lambda'_{\Delta 21})/2, \\
	\lambda^q_{\Phi 1\Delta 12} \equiv {}& (\lambda_{\Phi 1\Delta 12} + \lambda'_{\Phi 1\Delta 12})/2, \\
	\lambda^q_{1212} \equiv {}& \lambda_{1212} + \lambda'_{1212} \\
	\lambda^q_{2111} \equiv {}& (\lambda_{2111} + \lambda_{1211})/2, \\
	\lambda^{q\prime}_{2111} \equiv {}& (\lambda'_{2111} + \lambda'_{1211})/2, \\
	\lambda^q_{1222} \equiv {}& (\lambda_{1222} + \lambda_{2122})/2, \\
	\lambda^{q\prime}_{1222} \equiv {}& (\lambda'_{1222} + \lambda'_{2122})/2.
\end{align}
From these conditions we can rewrite $m^2_{\Phi 1}$, $M^2_1$ and $M^2_2$ as
\begin{align}
    m^2_{\Phi 1} = {}
		& \lambda_{\Phi 1} v^2
		+ u^2 \left(
			\lambda^q_{\Phi 1 \Delta 1} \cos^2\beta
			+ \lambda^q_{\Phi 1\Delta 2} \sin^2\beta
			+ 2 \lambda^q_{\Phi 1 \Delta 12} \cos\beta \sin\beta
		\right) \nonumber\\
		& - \sqrt{2} u \left(
			\mu_{11} \cos\beta
			+ \mu_{21} \sin\beta
		\right) \\
    M^2_1 = {}
		& - u^2 \cos^2\beta \left[
			\lambda^q_1
            + 3 \left(\lambda^q_{2111} + \lambda^{q\prime}_{2111}\right) \tan\beta
            + (\lambda^q_{12} + \lambda^q_{1212}) \tan^2\beta
			+ \left(\lambda^q_{1222} + \lambda^{q\prime}_{1222}\right) \tan^3\beta
        \right] \nonumber\\
		&
			- v^2 \left(
				\lambda^q_{\Phi 1\Delta 1}
				+ \lambda^q_{\Phi 1\Delta 12} \tan\beta
			\right)
			- M^2_{12} \tan\beta
			+ \frac{\mu_{11} v^2}{\sqrt{2}u \cos\beta}
			\\
    M^2_2 = {}
		&
			- u^2 \sin^2\beta \left[
				\lambda^q_2
				+ 3 \left(\lambda^q_{1222} + \lambda^{q\prime}_{1222}\right) \cot\beta
				+ (\lambda^q_{12} + \lambda^q_{1212}) \cot^2\beta
				+ \left(\lambda^q_{2111} + \lambda^{q\prime}_{2111}\right) \cot^3\beta
			\right]
		\nonumber\\
		&
			- v^2 \left(
				\lambda^q_{\Phi 1\Delta 2}
				+ \lambda^q_{\Phi 1\Delta 12} \cot\beta
			\right)
			- M^2_{12} \cot\beta
			+ \frac{\mu_{21} v^2}{\sqrt{2}u \sin\beta}
\end{align}

\subsection{Neutral states masses}

Consider the following expansion of the neutral states:
\begin{equation}
	\delta^0_n = \frac{\rho_n + i\eta_n}{\sqrt{2}}.
\end{equation}
We can take the base of the neutral states as
$S^0_\mathrm{even} = (h_1, \rho_1, \rho_2)$.
In such a basis, we have a $3\times 3$ mass matrix for the $CP$-even neutral scalars,
$M^2_\mathrm{even}$, which is symmetric and has elements given by:
\begin{align}
	\left(M^2_\mathrm{even}\right)_{11} = {}
		& 2 \lambda_{\Phi 1} v^{2} \\
	\left(M^2_\mathrm{even}\right)_{12} = {}
		& v \left[
			2 u \left(\lambda^q_{\Phi 1\Delta1} \cos\beta + \lambda^q_{\Phi 1\Delta12} \sin\beta\right)
			- \sqrt{2} \mu_{11}\right
		] \\
	\left(M^2_\mathrm{even}\right)_{13} = {}
		& v \left[
			2 u \left(\lambda^q_{\Phi 1\Delta2} \sin\beta + \lambda^q_{\Phi 1\Delta12} \cos\beta\right)
			- \sqrt{2} \mu_{21}\right]\\
	\left(M^2_\mathrm{even}\right)_{22} = {}
		&
		u^2 \cos^2\beta \left[
			2 \lambda^q_{1}
			+ 3 \left(\lambda^q_{2111} + \lambda^{q\prime}_{2111}\right) \tan\beta
			- \left(\lambda^q_{1222} + \lambda^{q\prime}_{1222}\right) \tan^3\beta
		\right]
		\nonumber\\
		&
		+ \frac{\mu_{11} v^2}{\sqrt{2} u \cos\beta}
		- (M^2_{12} + \lambda^q_{\Phi 1\Delta12} v^2) \tan\beta
		\\
	\left(M^2_\mathrm{even}\right)_{23} = {}
		& u^2 \cos\beta \sin\beta \left[
			2 (\lambda^q_{12} + \lambda^q_{1212})
			+ 3 \left(\lambda^q_{1222} + \lambda^{q\prime}_{1222}\right) \tan\beta
			+ 3 \left(\lambda^q_{2111} + \lambda^{q\prime}_{2111}\right) \cot\beta\right]\nonumber\\
		& + M^2_{12} + \lambda^q_{\Phi 1\Delta12} v^2 \\
	\left(M^2_\mathrm{even}\right)_{33} = {}
		&
			u^2 \sin^2\beta \left[
				2 \lambda^q_{2}
				+ 3 \left(\lambda^q_{1222} + \lambda^{q\prime}_{1222}\right) \cot\beta
				- \left(\lambda^q_{2111} + \lambda^{q\prime}_{2111}\right) \cot^3\beta
			\right]
		\nonumber\\
		&
			+ \frac{\mu_{21} v^2}{\sqrt{2} u \sin\beta}
			- (M^2_{12} + \lambda^q_{\Phi 1\Delta12}) \cot\beta.
\end{align}
For the $CP$-odd states we can take the basis
$S^0_\mathrm{odd} = (\eta_1, \eta_2)$,
obtaining the mass matrix elements
\begin{align}
\label{eq:m2odd11}
	\left(M^2_\mathrm{odd}\right)_{11} = {}
		&
		- u^2 \sin^2\beta \left[
			2 \lambda^q_{1212}
			+ \left(\lambda^q_{1222} + \lambda^{q\prime}_{1222}\right) \tan\beta
			+ \left(\lambda^q_{2111} + \lambda^{q\prime}_{2111}\right) \cot\beta
		\right]
		\nonumber\\
		&
		+ \frac{\mu_{11} v^2}{\sqrt{2} u \cos\beta}
		- (M^2_{12} + \lambda^q_{\Phi 1\Delta12} v^2) \tan\beta
		\\
\label{eq:m2odd12}
	\left(M^2_\mathrm{odd}\right)_{12} = {}
		&
		u^2 \cos\beta \sin\beta \left[
			2 \lambda^q_{1212}
			+ \left(\lambda^q_{1222} + \lambda^{q\prime}_{1222}\right) \tan\beta
			+ \left(\lambda^q_{2111} + \lambda^{q\prime}_{2111}\right) \cot\beta
		\right]
		\nonumber\\
		&
		+ M^2_{12}
		+ \lambda^q_{\Phi 1\Delta12} v^2\\
\label{eq:m2odd22}
	\left(M^2_\mathrm{odd}\right)_{22} = {}
		&
		- u^2 \cos^2\beta \left[
			2 \lambda^q_{1212}
			+ \left(\lambda^q_{1222} + \lambda^{q\prime}_{1222}\right) \tan\beta
			+ \left(\lambda^q_{2111} + \lambda^{q\prime}_{2111}\right) \cot\beta
		\right]
		\nonumber\\
		&
		+  \frac{\mu_{21} v^2}{\sqrt{2} u \sin\beta}
		- (M^2_{12} + \lambda^q_{\Phi 1\Delta12} v^2) \cot\beta
		.
\end{align}

The rest of the neutral scalars do not mix and the squared masses are given by the expressions
\begin{align}
m_{H^0}^2 = {}
	& u^2 \left[
		\left(\lambda_{\Phi 2\Delta1} + \lambda'_{\Phi 2\Delta1}\right) \frac{\cos^2\beta}{2}
		+ \left(\lambda_{\Phi 2\Delta2} + \lambda'_{\Phi 2\Delta2}\right) \frac{ \sin^2\beta}{2}
		+ \left(\lambda_{\Phi 2\Delta12} + \lambda'_{\Phi 2\Delta12}\right) \cos\beta \sin\beta
	\right] \nonumber\\
	& - \sqrt{2} u (\mu_{12} \cos\beta + \mu_{22} \sin\beta)
	+ m^2_{\Phi 2} + \frac{v^2}{2}\left(\lambda_{\Phi 12}  + \lambda'_{\Phi 12} + \lambda_5\right)\\
m_{A^0}^2 = {}
	& u^2 \left[
		\left(\lambda_{\Phi 2\Delta1} + \lambda'_{\Phi 2\Delta1}\right) \frac{\cos^2\beta}{2}
		+ \left(\lambda_{\Phi 2\Delta2} + \lambda'_{\Phi 2\Delta2}\right) \frac{ \sin^2\beta}{2}
		+ \left(\lambda_{\Phi 2\Delta12} + \lambda'_{\Phi 2\Delta12}\right) \cos\beta \sin\beta
	\right] \nonumber\\
	& + \sqrt{2} u (\mu_{12} \cos\beta + \mu_{22} \sin\beta)
	+ m^2_{\Phi 2} + \frac{v^2}{2}\left(\lambda_{\Phi 12}  + \lambda'_{\Phi 12} - \lambda_5\right)
\end{align}
In this case, the difference between the masses of these two scalars is given by
\begin{equation}
m_{A^0}^2 - m_{H^0}^2 = - \lambda_5 v^2 + 2\sqrt{2} u (\mu_{12} \cos\beta + \mu_{22} \sin\beta)
\end{equation}
which is positive (negative) when $H_0$ ($A_0$) is the DM candidate.

\subsection{Charged states masses}

In the scalar potential we have
three single charged scalars, $H^\pm$, $\delta_1^\pm$ and $\delta_2^\pm$;
and two doubly charged scalars, $\delta_1^{\pm\pm}$ and $\delta_2^{\pm\pm}$.
The usual charged Higgs present in two Higgs doublet models
does not mix with other charged scalars
and its squared mass is given by
\begin{equation}
	M_{H^\pm}^2 =
		 m_{\Phi 2}^2
		+ \frac{\lambda_{\Phi 12}}{2} v^2
		+ u^2 \sin\beta \cos\beta \left(
			\lambda_{\Phi 2\Delta 12}
			+ \frac{\lambda_{\Phi 2\Delta 1}}{2} \cot\beta
			+ \frac{\lambda_{\Phi 2\Delta 2}}{2} \tan\beta
		\right)\,.
\end{equation}
Note that the terms additional to the mass in the original IDM
have a factor of $u^2$.
Since we expect $u^2 \ll v^2$ these extra terms can be considered small corrections.
The two single-charge scalars from the triplets mix
and the $2\times 2$ matrix has the followings elements
\begin{align}
\label{eq:m2scharge11}
	\left(M^2_{\delta^\pm}\right)_{11} =
		{}&
		u^2 \sin^2\beta \left[
			\frac{\lambda'_{\Delta 12}}{4}
			- \frac{\lambda'_{\Delta 21}}{4}
			- \frac{\lambda^q_{12}}{2}
			- \lambda^q_{1212}
			- \left(\lambda^q_{1222} + \lambda^{q\prime}_{1222}\right) \tan\beta
			- \left(\lambda^q_{2111} + \lambda^{q\prime}_{2111}\right) \cot\beta
		\right]
		\nonumber\\
		&
		- M^2_{12} \tan\beta
		- v^2 \left(
			\lambda^q_{\Phi 1\Delta 12} \tan\beta
			+ \frac{\lambda'_{\Phi 1\Delta 1}}{4}
		\right)
		+ \frac{\mu_{11} v^2}{\sqrt{2} u \cos\beta}\,,
		\\
\label{eq:m2scharge12}
	\left(M^2_{\delta^\pm}\right)_{12} =
		{}&
		u^2 \sin\beta\cos\beta \left(
		- \frac{\lambda'_{\Delta 12}}{4}
		+ \frac{\lambda'_{\Delta 21}}{4}
		+ \frac{\lambda^q_{12}}{2}
		+ \lambda^q_{1212}
		+ \left(\lambda^q_{1222} + \lambda^{q\prime}_{1222}\right) \tan\beta
		+ \left(\lambda^q_{2111} + \lambda^{q\prime}_{2111}\right) \cot\beta
		\right)
		\nonumber\\
		&
		+M^2_{12}
		+ v^2 \left(
		\lambda^q_{\Phi 1\Delta 12}
		- \frac{\lambda'_{\Phi 1\Delta 12}}{4}
		\right)\,,
		\\
\label{eq:m2scharge22}
	\left(M^2_{\delta^\pm}\right)_{22} =
		{}&
		u^2 \cos^2\beta \left(
			\frac{\lambda'_{\Delta 12}}{4}
			- \frac{\lambda'_{\Delta 21}}{4}
			- \frac{\lambda^q_{12}}{2}
			- \lambda^q_{1212}
			- \left(\lambda^q_{1222} + \lambda^{q\prime}_{1222}\right) \tan\beta
			- \left(\lambda^q_{2111} + \lambda^{q\prime}_{2111}\right) \cot\beta
		\right)
		\nonumber\\
		&
		-M^2_{12} \cot\beta
		- v^2 \left(
			\lambda^q_{\Phi 1\Delta12} \cot\beta
			+ \frac{\lambda'_{\Phi 1\Delta 2}}{4}
		\right)
		+ \frac{\mu_{21} v^2}{\sqrt{2} u \sin\beta} \,.
\end{align}

The doubly charged scalars mix with each other
resulting in the following matrix elements
\begin{align}
\label{eq:m2dcharge11}
	\left(M^2_{\delta^{\pm\pm}}\right)_{11} =
		{}&
		u^2 \cos^2\beta\bigg[
			\lambda'_{\Delta 1}
			- \lambda^q_1
			- \tan\beta \left(
				3 \lambda^q_{2111}
				+ \lambda^{q\prime}_{2111}
			\right)
			- \tan^2\beta \left(
				\lambda^q_{12}
				+ \lambda^q_{1212}
				- \frac{\lambda'_{\Delta 12}}{2}
			\right)
		\nonumber\\
		&
			- \tan^3\beta \left(
				\lambda^q_{1222}
				+ \lambda^{q\prime}_{1222}
			\right)
		\bigg]
		+ v^2 \left(
			\frac{\mu_{11}}{\sqrt{2} u \cos\beta}
			- \frac{\lambda'_{\Phi 1\Delta 1}}{2}
			- \lambda^q_{\Phi 1\Delta 12} \tan\beta
		\right)
		\nonumber\\
		&
		- M^2_{12} \tan\beta
		\\
\label{eq:m2dcharge12}
	\left(M^2_{\delta^{\pm\pm}}\right)_{12} =
		{}&
		u^2 \sin\beta \cos\beta \left[
			\frac{\lambda'_{\Delta 21}}{2}
			+ \lambda'_{1212}
			+ \lambda^{q\prime}_{1222} \tan\beta
			+ \lambda^{q\prime}_{2111} \cot\beta
		\right]
		\nonumber\\
		&
		+ M^2_{12}
		+ v^2 \left(
			\lambda^q_{\Phi 1\Delta12}
			- \frac{\lambda'_{\Phi 1\Delta12}}{2}
		\right)
		\\
\label{eq:m2dcharge22}
	\left(M^2_{\delta^{\pm\pm}}\right)_{22} =
		{}&
		u^2 \sin^2\beta \bigg[
			 \lambda'_{\Delta2}
			- \lambda^q_{2}
			- \cot\beta \left(
			3 \lambda^q_{1222}
			+ \lambda^{q\prime}_{1222}
			\right)
			- \cot^2\beta \left(
				\lambda^q_{12}
				+ \lambda^q_{1212}
				- \frac{\lambda'_{\Delta12}}{2}
			\right)
		\nonumber\\
		&
			-  \cot^3\beta\left(
				\lambda^q_{2111}
				+ \lambda^{q\prime}_{2111}
			\right)
		\bigg]
		+ v^2 \left(
			\frac{\mu_{21}}{\sqrt{2} u \sin\beta}
			- \frac{\lambda'_{\Phi 1\Delta 2}}{2}
			- \lambda^q_{\Phi 1\Delta 12} \cot\beta
		\right)
		\nonumber\\
		&
		- M^2_{12} \cot\beta
\end{align}

\subsection{The no-mixing limit}

In the case where $u \ll v$
and assuming that the couplings in $V_{SB}$
are smaller than other couplings in the potential,
the non diagonal terms in mass matrices shown in this appendix
become subleading contributions.
We can call this \emph{the no-mixing limit}.
A small value for $u$ can be justified
by the need of having small neutrino masses
while small couplings in $V_{SB}$ can be
considered due to naturalness~\cite{tHooft:1979rat}.
The leading contributions to the diagonal elements are
\begin{align}
    \left(M^2_\mathrm{even}\right)_{11}& = 2\lambda_{\Phi 1} v^2 \\
    \label{eq:mdelta1nomix}
    \left(M^2_\mathrm{even}\right)_{22}& =
        \left(M^2_\mathrm{odd}\right)_{11} =
        \left(M^2_{\delta^{\pm}}\right)_{11} =
        \left(M^2_{\delta^{\pm\pm}}\right)_{11} =
        \frac{\mu_{11} v^2}{\sqrt{2} u \cos\beta} \\
    \label{eq:mdelta2nomix}
    \left(M^2_\mathrm{even}\right)_{33}& =
        \left(M^2_\mathrm{odd}\right)_{22} =
        \left(M^2_{\delta^{\pm}}\right)_{22} =
        \left(M^2_{\delta^{\pm\pm}}\right)_{22} =
        \frac{\mu_{21} v^2}{\sqrt{2} u \sin\beta},
\end{align}
With all the diagonal elements suppressed either by $u$ or couplings from $V_{SB}$.
The squared mass of the SM-like Higgs, $m_h^2$, can be identified with $2\lambda_{\Phi 1} v^2$.
In the case of the triplets, all the masses of the fields in each triplet become
degenerated and we can write $M_{\Delta 1}^2 = \mu_{11} v^2/\sqrt{2} u \cos\beta$
and $M_{\Delta 2}^2 = \mu_{21} v^2/\sqrt{2} u \sin\beta$,
for all the scalars contained in the triplets.
In the case of the masses for the dark scalars, $H_0$, $A_0$ and $H^{\pm}$,
it is easy to see that their masses fallback to the values in the original IDM.

\end{document}